\documentclass{aastex631}

\shorttitle{
Color and Size distributions of Jupiter Trojans}
\shortauthors{Yoshida et al.}
\usepackage{graphicx}
\usepackage{color}
\usepackage{url}

\begin{document}

\title{
Color and Size distributions of small Jupiter Trojans 
}

\correspondingauthor{Fumi Yoshida}
\email{fumi-yoshida@med.uoeh-u.ac.jp, fumi.yoshida@perc.it-chiba.ac.jp}

\author[0000-0002-3286-911X]{Fumi Yoshida}
\affiliation{University of Occupational and Environmental Health, Japan, 1-1 Iseigaoka, Yahata, Kitakyusyu 807-8555, Japan}
\affiliation{Planetary Exploration Research Center, Chiba Institute of Technology, 2-17-1 Tsudanuma, Narashino, Chiba, 275-0016, Japan}
\author[0000-0003-4143-4246]{Tsuyoshi Terai}
\affiliation{Subaru Telescope, National Astronomical Observatory of Japan, 650 North A`ohoku Place, Hilo, HI 96720, USA}
\author[0000-0002-4383-8247]{Keiji Ohtsuki}
\affiliation{Department of Planetology, Kobe University, Kobe 657-8501, Japan}



\begin{abstract}
We conducted a two-band imaging survey observation using the Subaru Telescope and its wide-field camera, Suprime-Cam, to study the visible colors and size distribution of Jupiter's Trojan asteroids.
The survey covered an area around Jupiter's L4 Lagrange point totaling 9.2 square degrees. We detected 120 Trojan asteroids in this survey.
From these Trojan asteroids, we extracted 44 unbiased samples with absolute magnitudes in the $g^\prime$ band ranging from 12.9 to 16.9~mag (corresponding to diameter ranges of approximately $\sim$3--16~km assuming an albedo of 0.05) and analyzed their $g^\prime - i^\prime$ color and size distributions.
%
Large Jupiter Trojan asteroids are known to be classified into two color groups, ``red'' and ``less red''.
We found that such bimodality in the color distribution is absent for small Jupiter Trojan asteroids, which is consistent with previous studies.
Previous studies have 
also shown that these two groups have different slopes in the magnitude distributions from each other, which was explained by conversion of red objects to less-red fragments through catastrophic disruptions.
In contrast, we found that the size frequency distributions of our two sample groups divided by the color of $g^\prime - i^\prime$~=~0.7 (in AB magnitude) are quite similar.
Our results can provide new insights into collisional evolution of color and size distribution of small Jupiter Trojans.
%
\end{abstract}



\section{Introduction} \label{sec:intro}
The Jupiter Trojans (hereafter referred to as the JTs) are a group of asteroids that orbit around the Lagrangian points L4 and L5, which are located 60 degrees ahead and behind Jupiter, respectively.
These asteroids librate around the Lagrangian points, making up swarms that are about 70–80 degrees wide along Jupiter's orbit.
Previous estimates have suggested that the total number of asteroids in the L4 and L5 swarms is comparable to that of the main-belt asteroids \citep{jewitt2000,szabo2007,NY2008}.
However, recent surveys for small JTs \citep{yoshida2017,uehata2022} and main-belt asteroids \citep{Maeda_2021} using the Subaru Telescope and its wide-field prime focus camera Hyper Suprime-Cam \citep{miyazaki2018} suggest that the number of JTs larger than 1~km in diameter ($D$) is about $2.6 \times 10^{5}$ (including both L4 and L5 populations), while that of main belt asteroids is $\sim2 \times 10^{6}$. Thus, the total number of JTs with $D > 1$~km is significantly smaller than that of main belt asteroids in the same size range (perhaps by almost one order of magnitude). 
Previous surveys showed that the L4 and L5 swarms are asymmetrical in their total number, $N_\mathrm{L4}$ and $N_\mathrm{L5}$, with $N_\mathrm{L4}/N_\mathrm{L5} \sim 1.3 -1.9$ \citep{szabo2007,NY2008,grav2011,grav2012b}. 
Recent surveys with the Hyper Suprime-Cam of the Subaru Telescope also confirmed this asymmetry and showed $N_\mathrm{L4}/N_\mathrm{L5} \simeq 1.40 \pm 0.15$ for JTs with $D > 2$~km \citep{yoshida2017,uehata2022}, which is consistent with the ratio evaluated from the background (non-family) JT populations for the absolute magnitude $H < 15$, $1.45 \pm 0.05$, by \cite{Vokrouhlicky2024}.

Observational studies of JTs can provide unique constraints on their origin as well as the orbital evolution of giant planets. 
For example, if JTs originated from planetesimals in the vicinity of Jupiter's orbit and were captured in the Trojan regions during Jupiter’s formation period \citep{marzari1998}, they could reveal information about Jupiter's radial location and the composition of its solid core during its formation. This is possible even if Jupiter experienced significant orbital migration after its formation \citep{pirani2019}. 
On the other hand, if the JTs originated as planetesimals in the primitive Kuiper Belt that were captured in the Trojan regions during the instability of the giant planets' orbits, they could provide insight into the orbital evolution of the giant planets and small bodies in the outer Solar System \citep{morbidelli2005,nesvorny2013}.
Thus, the colors of the JTs, especially in relation to those of other small-body populations such as Kuiper-belt objects (KBOs), would give us important clues about their formation location and subsequent orbital evolution \citep{Jewitt2018}.

The size distribution of JTs has been studied in detail in previous studies \citep[e.g.,][]{jewitt2000,szabo2007,yoshida2005,yoshida2008,fraser2014,wong2014,wong2015,yoshida2017,Vokrouhlicky2024}. 
These studies have provided knowledge of the size-frequency distribution of JTs down to 2 km in diameter.
\cite{yoshida2019,yoshida2020} compared the size-frequency distribution of JTs with those of main-belt asteroids, Hilda asteroids, and the craters on Pluto and Charon (reflecting the size-frequency distributions of small KBOs) studied by the New Horizons mission, and found a gradual transition of the size-frequency distributions from the inner region (i.e., outer main asteroid belt) to the outer region (i.e., the Jupiter Trojan).
\cite{yoshida2019,yoshida2020} argued that this gradual transition could be explained if some portion of small KBOs were mixed with JTs, Hildas, and outer main-belt asteroids in different proportions.
This interpretation is consistent with a model in which the present-day JTs are captured objects from the Kuiper belt during planetary migration.
In addition, \cite{uehata2022} found that the size distribution of JTs with $D > 2$~km is indistinguishable between the L4 and L5 swarms over a wide size range, which supports the idea that the two populations have the same origin and experienced similar collisional evolutions.
\cite{terai2018} showed that the size distribution of the JTs also coincides with that of Hilda asteroids in the range of $D \sim$~1--10~km.
This suggests that Hildas and JTs in this size range likely underwent similar collisional evolutions and, possibly, share a common origin.

The current knowledge of the surface properties of JTs is reviewed in \cite{Emery2024}, and their main characteristics can be summarized as follows:
(1) JTs are a more uniform population than main belt asteroids \citep[e.g.,][]{Dotto2006420}.
(2) The average albedo of JTs is low \citep[e.g.,][]{grav2011}.
(3) There are two groups classified with visible and near-infrared spectra: one group is red (corresponding to D-type asteroids) and the other group is less red/neutral (corresponding mostly to P-type asteroids, but also including C-type asteroids) \citep[]{szabo2007,roig2008,szabo2007,roig2008}.
According to \cite{demeo2013}, approximately 67\% of JTs are D-type and 30\% are X-type (mostly P-type) asteroids.
(4) No clear features were found in the visible and near-infrared spectra \citep{dotto2006,fornasier2007,yang2007,melita2008}.

Regarding the two groups of spectra mentioned in (3) above, more recent research findings suggest that a third group of high-albedo JTs exist whose spectra differ from those of the red and less red subgroups \citep{Brown_2025}. Absolute magnitude measurements by the Zwicky Transient Facility (ZTF) and 870 $\mu $m thermal emission observations by the Atacama Large Millimeter/submillimeter Array (ALMA), which targeted midsized JTs (15--40 km in diameter), revealed that, while the median albedo of typical JTs is 0.05 -- 0.07 \citep{grav2012b}, some JTs have significantly high albedo of 0.09 -- 0.13 \citep{Simpson_2022}. 
Detailed spectral observations of high-albedo JTs in the 0.8 -- 5 $\mu $m range, conducted by \cite{Brown_2025} using the James Webb Space Telescope (JWST), revealed that the spectral characteristics of high-albedo JTs differ distinctly from those of the red and less red groups in the slope from the visible to the near-infrared range and at wavelengths greater than 3 $\mu $m (see Figure 5 in \cite{Brown_2025}). The authors propose defining high-albedo JTs as a third group, distinct from the red and less red groups. Interestingly, this third group consists of relatively small JTs (approximately 20 km in diameter) that may potentially reflect consequences related to surface impact evolution.

The most enigmatic unsolved question regarding JT groups is as follows: Generally, while different colors/spectra of objects indicate different origins or environments, “why do red and reddish/neutral objects coexist in such localized regions?”
%
To address this question, we considered it necessary to first examine the colors of small JTs.
Since small JTs are likely fragments of larger parent bodies, their surface composition may reflect that of the parent body's interior at least to some extent.
Therefore, observing the surface composition of small JTs could provide clues about the formation location or origin of the parent body.
We used the 8.2-m Subaru Telescope.
Its large aperture allows us to detect small JTs down to $D = 2$~km.

Previous studies have also examined this subject. 
Using the Sloan Digital Sky Survey (SDSS) Moving Object Catalog (MOC) 4, \cite{wong2014} classified 221 JTs with absolute magnitudes in the range of $H < 12.3$ into two color populations, ``red'' and ``less red''.
\cite{wong2015} conducted a survey of L4 JTs in the $g^\prime$- and $i^\prime$-bands using the Subaru Telescope and Suprime-Cam \citep{miyazaki2002} to study the relationship between the color and size frequency distributions of small JTs.
%
The survey covered around the L4 swarm (38 square degrees in total).  
They detected 557~JTs with $H$~=~10.0$-$20.3~mag. 
Due to poor seeing conditions on the first night of their observation, they chose a seeing cutoff of $1$\farcs$2$ to define their sample, and used 150 out of the 557~JTs for color analysis. 
Each field was visited four times with a cadence of $g^\prime~\rightarrow~g^\prime~\rightarrow~i^\prime~\rightarrow~i^\prime$ bands. 
The filters were changed only once during the four visits for each survey field.
The imaging interval was 20~minutes in each band and 30~minutes when changing the filters from $g^\prime$ to $i^\prime$ or vice versa.
The exposure time was 60~seconds for each imaging. 
The total time to complete imaging of each field was approximately 74~minutes.
With this imaging method, the average magnitudes of the $g^\prime$ and $i^\prime$ bands were obtained approximately 50~minutes apart. Thus, the effect of brightness variation due to the rotation of each object seems to be not negligible. 
%
From this observation, \cite{wong2015} did not find bimodality in the color distribution, and explained that this is due to the blurring effect of asteroid rotation.
\cite{pan2022} examined the $g-i$ and $g-r$ colors of JTs detected in the six-year data set of the Dark Energy Survey (DES), and obtained results similar to those of \cite{wong2015}. 
%
Although the sample size of JTs detected in the DES is large, the JTs were observed at different epochs in each band; thus, their measured  colors contain uncertainties due to the effects of light variations caused by rotation.

In our study, we used a different method in an attempt to reduce the effect of JT's rotation.
We made the observations described below.

\begin{figure}
\plotone{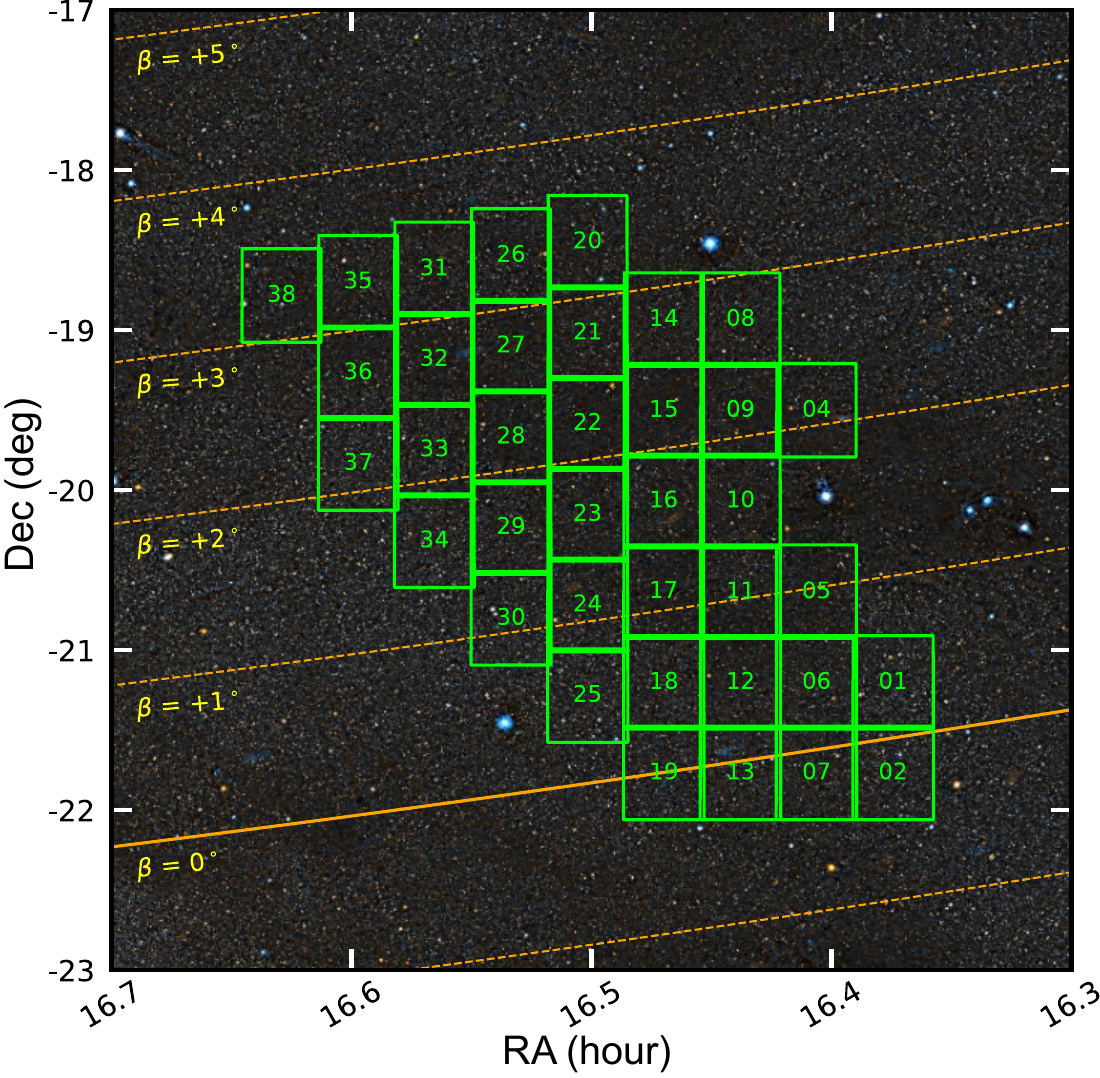}
\caption{Locations of our survey fields.
The size of each rectangle corresponds to the field of view of Suprime-Cam. 
The solid line is the ecliptic plane, and the dashed lines show ecliptic latitudes with an interval of 1$\arcdeg$.
The background image is from Pan-STARRS.
\label{fig:obsfield}}
\end{figure}

\section{Observation} \label{sec:obs}
We conducted imaging surveys of the L4 swarm of JTs using the $g^\prime$-band and $i^\prime$-band filters with the wide-field camera, Suprime-Cam, mounted on the 8.2-m Subaru Telescope at the summit of Maunakea on May~24 and 30, 2017~UT.
The Suprime-Cam is a mosaic CCD camera consisting of 10~2048~$\times$~4096 CCDs covering a $34\arcmin \times 27\arcmin$ field of view (FoV) corresponding to 0.25~deg$^2$ with a pixel scale of 0$\farcs$20 \citep{miyazaki2002}.

The Suprime-Cam is the first-generation instrument of the Subaru Telescope and has been at the forefront of research as a wide-field camera equipped on a large telescope for about 18~years.
It was then decommissioned in May 2017 and completely replaced by the second-generation ultra-wide-field camera, Hyper Suprime-Cam \citep{miyazaki2018}. Our observations were made during Suprime-Cam's final observing run. 
On the final night of the last run, May~30, 2017~UT, Suprime-Cam heavy users, developers, and support astronomers gathered to observe its final operation and show their appreciation for Suprime-Cam's 18~years of work. 
We were honored to use Suprime-Cam for our research in its final run.

Although the Suprime-Cam was an older unit, it had one significant advantage over the Hyper Supreme-Cam—the ability to change filters quickly.
When measuring an asteroid's brightness at different times with various filters to determine its colors, it is important to mitigate the effects of light variation caused by its rotation.
Therefore, Suprime-Cam's ability to quickly change filters is advantageous for color observations.

We selected the survey fields according to the following conditions: 
(1)~the distance from the L4 point is less than 20$\arcdeg$ in ecliptic longitude, 
(2)~the distance from the opposition on the observation day is less than 10$\arcdeg$ in ecliptic longitude,
(3)~the ecliptic latitude is within 5$\arcdeg$ from the ecliptic plane, 
(4)~the galactic latitude is larger than 18$\arcdeg$, and 
(5)~the survey area is more than 4$\arcdeg$ away from the nearby bright star Antares ($\alpha$~Sco).
The survey area covers 9.2~deg$^2$ centered at (RA, Dec) = (16:32:00, -20:00:00).
We surveyed 14~FoVs of Suprime-Cam on May~24, 2017 (UT) and 23~FoVs on May~30, 2017 (UT).
In total, 37~fields were observed using both the $g^\prime$-band and $i^\prime$-band filters (see Figure ~\ref{fig:obsfield}). 
Each field was visited three times in the order of $i^\prime~\rightarrow~g^\prime~\rightarrow~i^\prime$ at 20--30~min intervals, corresponding to an observation arc of 40--60~min.
The exposure time for a single image was 90~sec in each band. 
The seeing size range was from 0$\farcs$54 to 1$\farcs$5.
A summary of our observations is shown in Table~\ref{table:ObsLog} in Appendix \ref{appen:obslog}.

\section{Data Reduction and Analysis} \label{sec:ana}

We reduced our survey data taken by Suprime-Cam using the dedicated pipeline SDFRED2 \citep{ouchi2004}.
The images were processed on each CCD chip with the following procedure: 
(1)~bias subtraction and flat fielding, 
(2)~image distortion correction, (3)~sky subtraction, (4)~WCS correction, and (5)~photometric calibration using relatively bright objects (17--21~mag in the $i^\prime$~band) cross-matched with the Pan-STARRS~1 (PS1) Data Releases~2 (DR2) catalog \citep{flewelling2020}.

First, we used SExtractor \citep{bertin1996} to detect light sources in each processed image. We set the detection thresholds of 2.0 and 3.0 for the $g^\prime$-band and $i^\prime$-band data, respectively, and created a catalog of light sources.
Next, we searched for light source combinations as moving object candidates based on the rate and direction of sky motion using our detection code. This search was conducted across all light sources in the combined catalogs compiled from all data sets.
This procedure is the same as in \cite{uehata2022} to search for moving objects.
Then, we simulated the possible range of apparent sky motions of L4 JTs in our survey fields using a Monte Carlo method based on the orbital distribution of known JTs for each observation night. This method generated tens of thousands of synthetic objects. This corresponds to an arcuate range roughly from $-$28$\arcsec$~hr$^{-1}$ (although most of them are $> -$22$\arcsec$~hr$^{-1}$) to $-$17$\arcsec$~hr$^{-1}$ along the ecliptic longitude and within $\pm$16$\arcsec$~hr$^{-1}$ along the ecliptic latitude \citep[see Figure~2 of][as reference]{uehata2022}.
Among the detected moving objects, those within the estimated range of apparent sky motion were extracted and designated as JT candidates.
Each candidate was visually inspected in the image to determine whether it was a moving object. To aid in the visual confirmation, a diagram showing the motion and image of the JT candidates from three visits at a glance was created (see Figure~\ref{eye-inspection} in Appendix \ref{appen:ex_visualinspection}).
Ultimately, 120 objects were identified as JT asteroids.

We measured the fluxes of the detected JT asteroids using aperture photometry with the {\tt APPHOT} package in IRAF \citep{Tody_1986,Tody_1993}.
Due to the short exposure time relative to the sky motions of JT asteroids, JT asteroids appear almost indistinguishable from point sources in the image.
We used a circular aperture with a diameter of $2\farcs0$ for photometry.
If stars/galaxies overlapped on the aperture area, we subtracted the image containing the JT from a reference image obtained at the same field in a different exposure using the optimal image subtraction technique developed by \citet{bramich2008}. 
This allowed us to measure the flux of the JT without contamination from stars or galaxies.
However, photometry failed on 16 objects in the $g^\prime$-band images due to weak signals or unsuccessful subtraction of nearby background objects.
Ultimately, we obtained a $g^\prime - i^\prime$ color sample from 104~JTs.
We converted the $g^\prime - i^\prime$ color taken by Suprime-Cam's filters to the SDSS filter system on the AB magnitude scale to allow for comparison with previous studies.

The primary objective of this study is to investigate the relationship between the color and size distribution of JTs. 
To accomplish this, we must derive the absolute magnitude of the detected JTs from their apparent magnitude and heliocentric/geocentric distances.
We estimate the heliocentric distance of each asteroid from its sky motion, assuming a non-eccentric orbit using the expressions presented in \citet{terai2013}. 
We then corrected the estimate using a linear regression derived from a Monte Carlo simulation with thousands of synthetic JT orbits, as described in \citet{yoshida2017}, $R_{\rm corrected} = 2.05 R_{\rm original} - 5.44$, where $R_{\rm original}$ and $R_{\rm corrected}$ are the original and corrected heliocentric distances, respectively.
The random error of the resulting heliocentric distance is 0.11~au ($\sim$2\%).
In order to accurately analyze the relationship between absolute magnitude and color distributions, it is essential to: (1)~assess the magnitude-dependency of the detection efficiencies, and (2)~select an unbiased sample of the detected objects.
Regarding the first item, we used the same method as \citet{uehata2022} that a number of synthetic objects generated with an equivalent seeing size and luminosity as the actual data are implanted throughout the processed images.
We estimated the detection efficiency at 21~mag (bright enough to be used as a calibration for one of the two filters) and from 23~mag to 25~mag in 0.2~mag increments, as we did for JT detection with SExtractor.
We calculated a best-fit curve for the detection efficiencies in each exposure and for each CCD, using least-squares optimization with the following expression:
\begin{equation}
  \eta (m) = \eta_0 \Bigg( \frac{ \epsilon (m) }{ 2 } \left[ 1 - \tanh \left( \frac{ m^p - m_{50} } {w_1 } \right) \right]
                  + \frac{ 1 - \epsilon (m) }{ 2 } \left[ 1 - \tanh \left( \frac{ m - m_{50} }{ w_2 } \right) \right] \Bigg), 
\end{equation}
where $m$ is the apparent magnitude given to the implanted synthetic objects, $\eta_0$ is the peak value derived at 
$m$~=~21~mag, $m_{50}$ is the magnitude at $\eta (m) = \frac{ \eta_0 }{ 2 }$, $w_{\{1,2\}}$ are the transition widths, 
$p$ is the exponent in the power-law approximation of $\eta (m)$ at the bright-end range, and $\epsilon (m)$ is the fraction of each term given by 
\begin{equation}
  \epsilon (m) = \frac{ 1 }{ 2 } \left[ 1 - \tanh \left( \frac{ m - m_{50} }{ 0.2 } \right) \right].
\end{equation}

\begin{figure}[t!]
\plotone{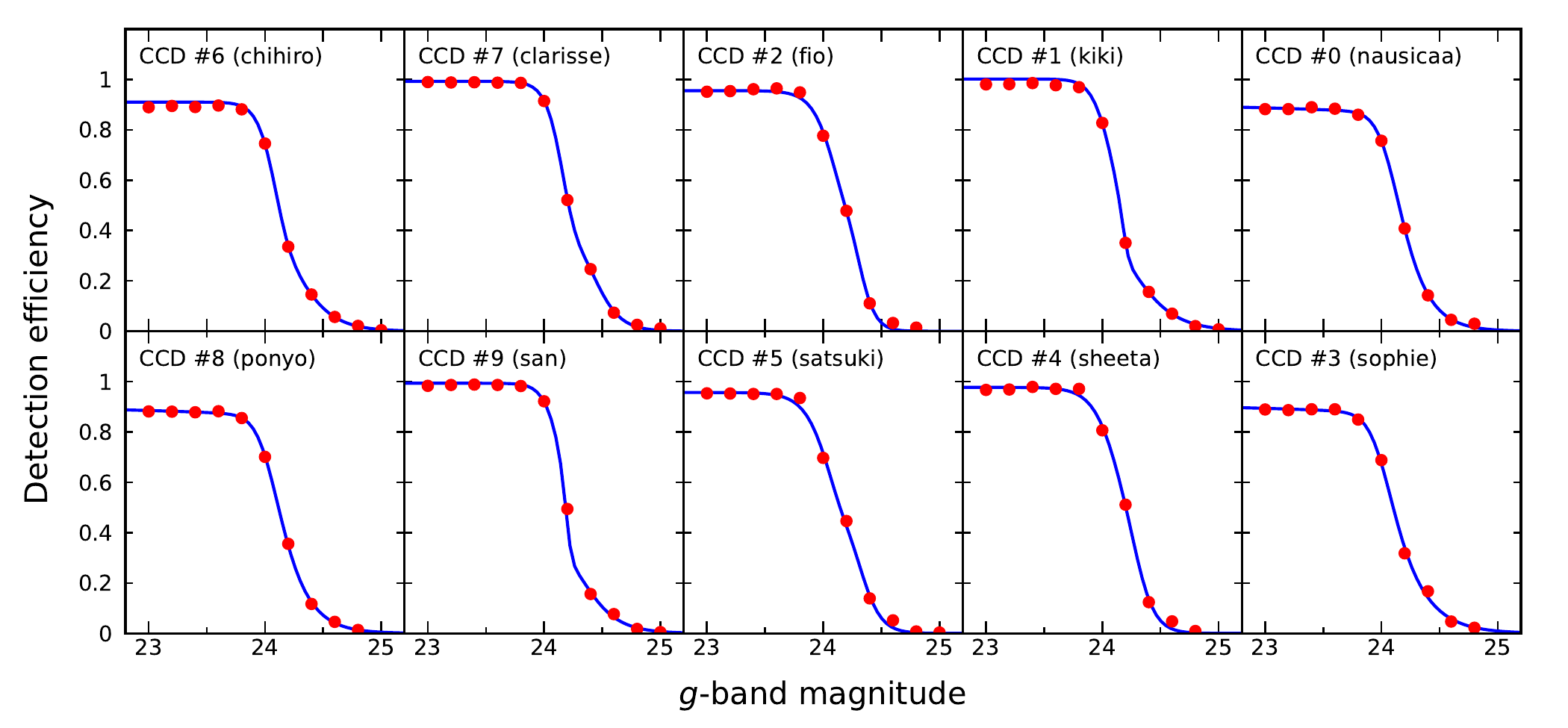}
\caption{
Examples of JT detection efficiencies measured from 90-s exposures taken with a $g^\prime$ band filter on May 24, 2017~UT, under sunny conditions $\sim$ 1$\farcs$0 seeing.
Each panel with the detector ID (in parentheses the name of the CCD chip) in the upper left corner corresponds to the actual layout of the CCD mosaic on the Suprime-Cam focal plane.
CCDs~\#0, \#3, \#6, and \#8 are located at the corners of the FoV of the Suprime-Cam.
The solid lines are best-fit curves (see text).
\label{fig:deteff}}
\end{figure}

Figure~\ref{fig:deteff} shows an example of the detection efficiency curves obtained from an exposure taken with the $g^\prime$~band under an average seeing condition of $\sim$1$\farcs$0.
It can be seen that the $\eta_0$ values on the CCDs~\#0, \#3, \#6, and \#8, which are located at the corners of the FoV of the Suprime-Cam are slightly lower than those on the other CCDs due to the effect of vignetting.
We confirmed that the $m_{50}$ values are higher than 23.9~mag in most of the $g^\prime$ band images, and defined the detection limit of our survey as 23.9~mag in the $g^\prime$ band.
Note that since the $g^\prime$ band is less sensitive to JTs than the $i^\prime$~band in our observations, we specified the detection limit using the former.

For the second item, we selected an unbiased sample based on the heliocentric distance and the absolute magnitude of the $g^\prime$ band, given by 
$H_g = m_g - 5 \log(R \cdot \Delta)$, where $R$ and $\Delta$ are the heliocentric and geocentric distances in au, respectively, and $m_g$ is the $g^\prime$-band apparent magnitude.
The uncertainty of 0.11~au in the corrected heliocentric distance mentioned above propagates to an error of 0.10~mag in $H_g$, which dominates the photometric error for most objects.

\begin{figure}[t!]
\plotone{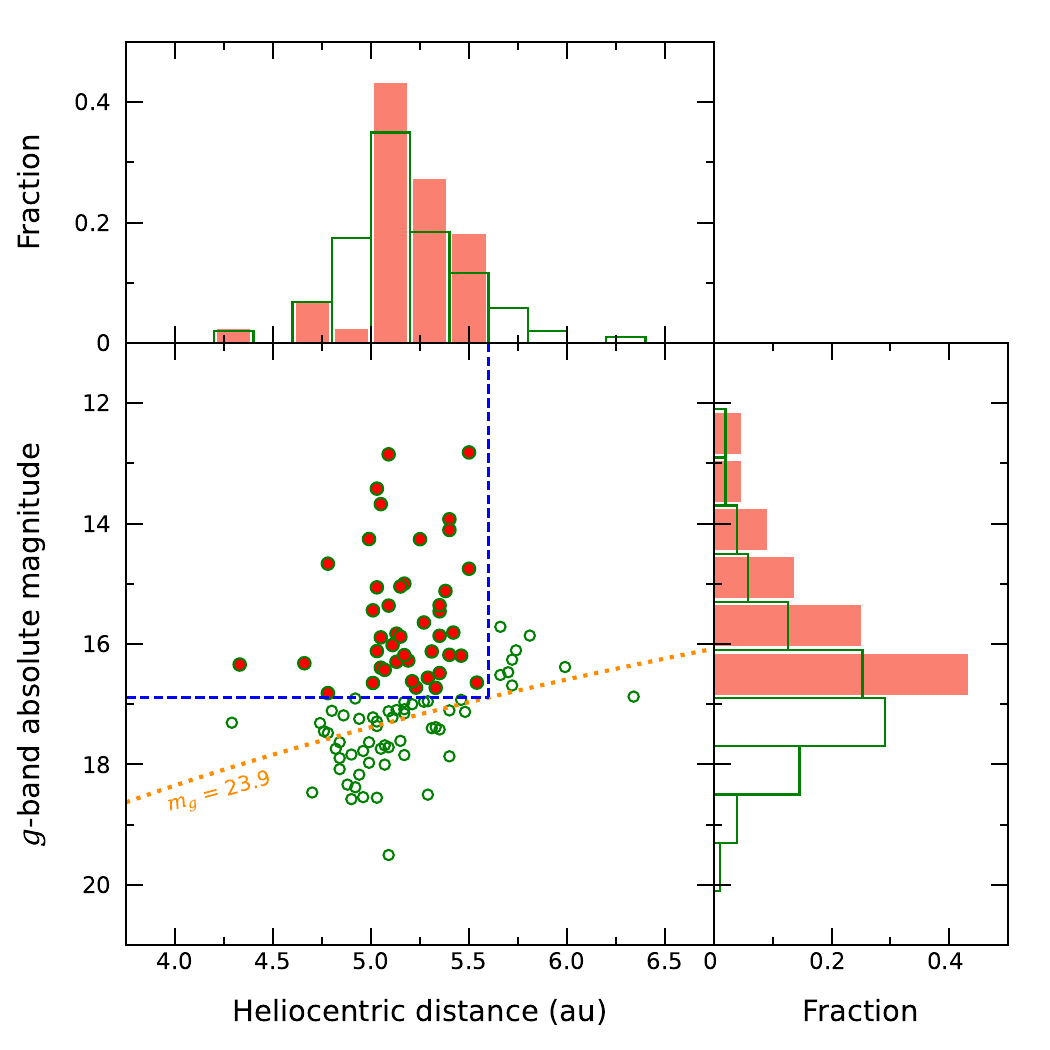}
\caption{
Lower left: The distribution of heliocentric distance ($R$) vs. $g^\prime$-band absolute magnitude ($H_g$) of the detected JT asteroids.
The uncertainty of the estimated heliocentric distances is $\sim$0.11~au, which causes the $H_g$ errors of $\sim$0.1~mag as the primary factor for most objects.
The filled circles are the objects with $R$~$<$~5.6~au and $H_g$~$<$~16.9~mag (separated by the dashed lines) defined as the unbiased sample.
The detection limit of our data, namely the apparent magnitude of $m_g$~=~23.9~mag, is indicated by the dotted line.
Upper left and lower right: the $R$ and $H_g$ histograms of the full sample (open bars) and the unbiased sample 
(solid bars), respectively.
\label{fig:r-H}}
\end{figure}

Figure~\ref{fig:r-H} shows the distribution of $H_g$ versus $R$ of the detected JTs as well as their histograms.
The upper limit of $R$ for object sampling is set to 5.6~au in order to maximize the number of samples.
This corresponds to $H_g$~=~16.9~mag at the detection limit of $m_g$~=~23.9~mag.
Accordingly, 44 objects were extracted from the range of $R <$~5.6~au and $H_g <$~16.9~mag as an unbiased bicolor JT sample.
Assuming a geometric albedo of 0.05 \citep{Romanishin_2018}, the $H_g$ limit is approximately equivalent to a diameter of 3~km.

\section{Results} \label{sec:results}
\begin{figure}[t!]
\plotone{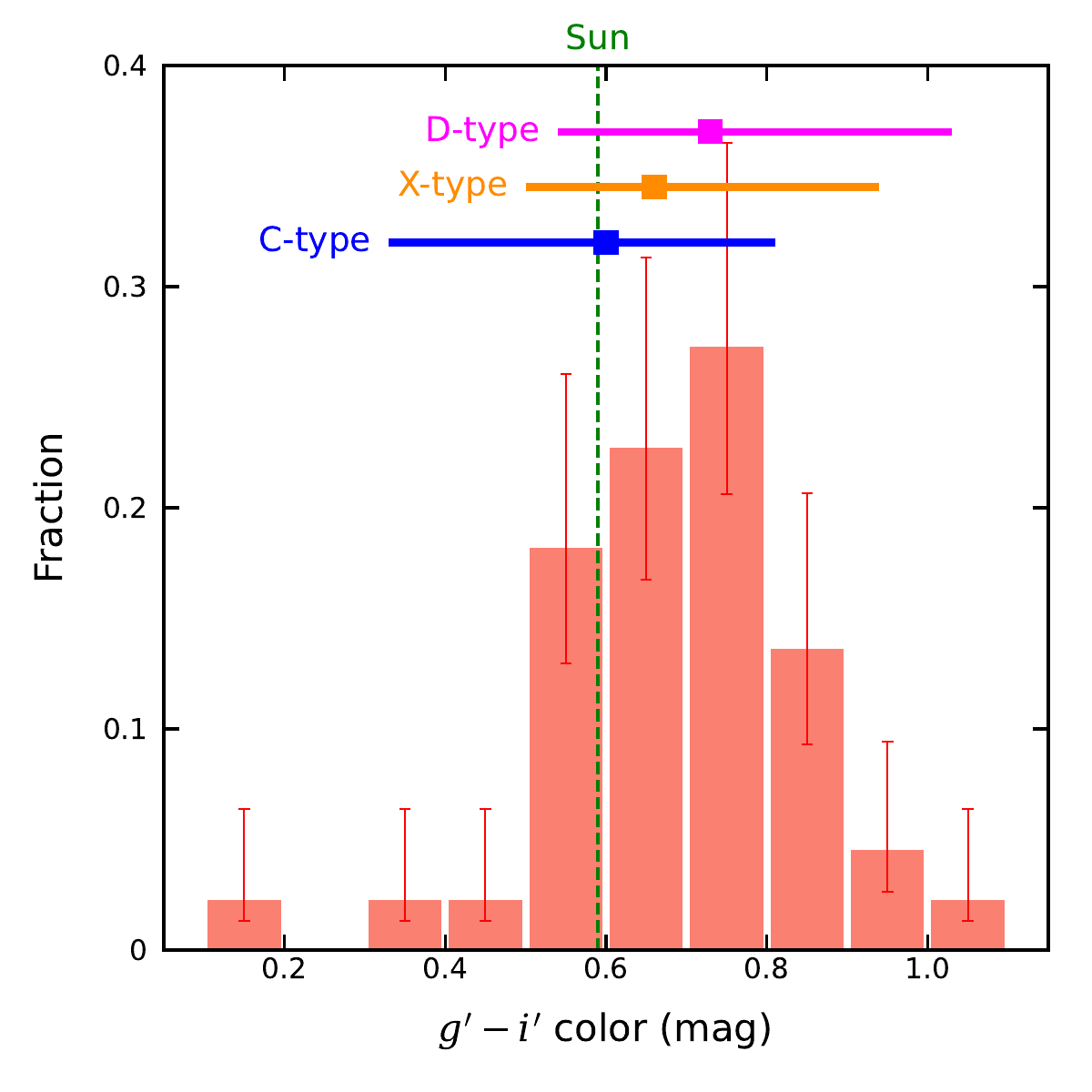}
\caption{
The histogram of the $g^\prime - i^\prime$ color distribution of the unbiased JT sample.
Error bars are based on Poisson errors.
The dashed line shows the color of the Sun \citep{bertin1996}.
The squares with solid lines show the mode values and color ranges, respectively, of the C-, X-, and D-type asteroids derived from the SDSS-based Asteroid Taxonomy database \citep{hasselmann2011}.
\label{fig:color}}
\end{figure}

Figure~\ref{fig:color} shows the histogram of the $g^\prime - i^\prime$ colors of our unbiased JT sample.
The color ranges of the D-, X- and C-types in Figure~\ref{fig:color} were computed by using the SDSS Taxonomy Templates\footnote{\url{https://sbn.psi.edu/pds/resource/sdsstax.html}} \citep[PDS4 version;][]{hasselmann2011} and the classification scheme developed by \citet{carvano2010}. 
The typical $g^\prime - i^\prime$ color ranges are calculated to be
$0.73^{+0.30}_{-0.19}$~mag for the D~type, $0.66^{+0.28}_{-0.16}$~mag for the X~type, and $0.60^{+0.21}_{-0.27}$~mag for the C~type, respectively.
The color distribution of our sample mostly overlaps with the D- and X-type ranges, which is consistent with the taxonomic distribution of known JTs.

%

Figure~\ref{fig:color} is analogous to the one derived from the Suprime-Cam data obtained by \citet{wong2015}.
%
However, the clear bimodality of color distribution identified by \cite{roig2008} and \cite{Emery_2011}, and also demonstrated by \cite{wong2014} using SDSS-MOC4, is not observed.
To divide the unbiased sample into two groups of the ``red'' and ``less red'' populations, we define the boundary color as $g^\prime - i^\prime$~=~0.70~mag, which is located halfway between the mean colors of the D-type (red) and X-type (less red) asteroids.
Coincidentally, the boundary color of $g^\prime - i^\prime$~=~0.70~mag is very close to the average $g^\prime - i^\prime$ of 0.69 $\pm$ 0.16 in the unbiased sample. This value is also identical to the average $g^\prime - i^\prime$ for the entire sample. However, the entire sample has a larger deviation ($\pm$0.43).
This boundary also corresponds exactly to the median of the unbiased sample; that is, it divides the red and less-red groups equally, with 22~objects in each group.

The size distribution of each group is evaluated as the distribution of the cumulative number of objects with $H_g$ brighter than the $j$-th ($j = 1, 2, \cdots, N_{\rm group}$, where $N_{\rm group}$ is the number of the objects classified as belonging to one of the two groups) brightest object that is corrected by the detection efficiency of each object.
The cumulative number is defined as $N (< H_g)$, expressed as
\begin{equation}
N (< H_g) = \sum_{ j: H_{g,j} < H_g } \Bigg[ \prod_{ k = 1 }^{ 3 } \eta_k (m_{j,k}) \Bigg]^{-1},
\end{equation}
where $H_{g,j}$ and $\eta_k (m_{j,k})$ represent $H_g$ of the $j$-th objects and $\eta (m)$ of the $k$-th exposure
(in the $g^\prime$ band if $k = 2$, otherwise in the $i^\prime$ band) for the $j$-th object, respectively.

\begin{figure}[t!]
\plotone{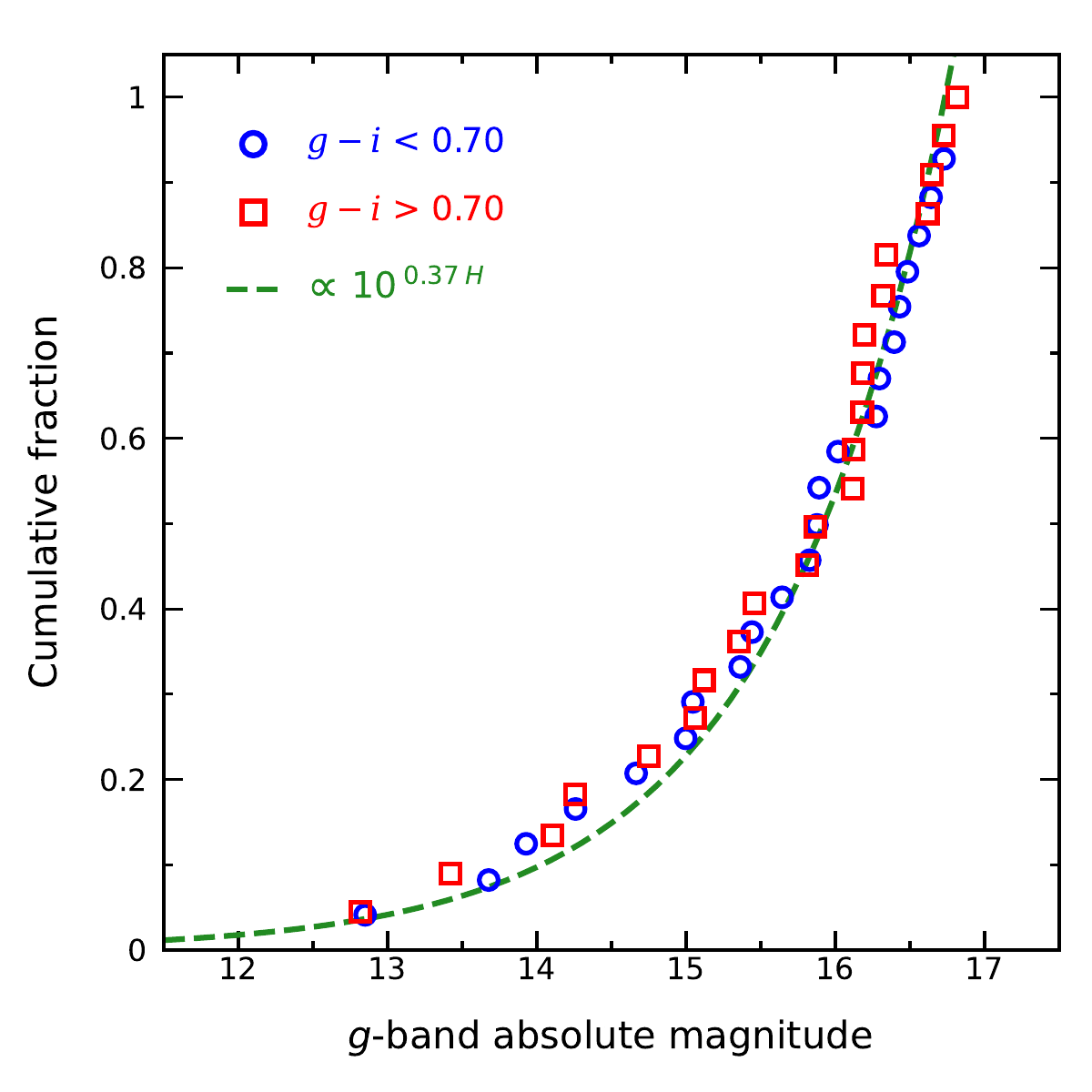}
\caption{
The normalized absolute magnitude distributions of the ``less red'' (circles) and ``red'' (squares) objects separated by $g^\prime -i^\prime$~=~0.70~mag. 
The dashed line shows the power-law function for the absolute magnitude distribution of L4 JTs obtained by \citet{yoshida2017}.
\label{fig:csds}}
\end{figure}

Figure~\ref{fig:csds} shows a direct comparison of the normalized $H_g$ distributions of the ``less red'' ($g^\prime - i^\prime$~$<$~0.70~mag) and ``red'' ($g^\prime - i^\prime$~$>$~0.70~mag) objects.
The two distributions appear quite similar.
Also, both agree well with a single-slope power law of $N (< H) \propto 10^{0.37H}$ reported by \citet{yoshida2017} as the best-fit function for the absolute magnitude distribution of 481 L4 JTs in a similar size range as our sample. This indicates that these $H_g$ distributions are accurately measured even with smaller sample sizes.

We used the two-sample Kolmogorov-Smirnov (K-S) test \citep{stephens1970, press1992} to 
quantitatively evaluate the difference between these two cumulative $H_g$ distributions, $S_{\rm R} (H_g)$ and $S_{\rm LR} (H_g)$ for the ``red'' and ``less red'' groups, respectively, with the null hypothesis that the two groups have identical distributions.
%
The K-S test yields a $p$-value of 0.98, indicating that the distributions of $S_{\rm R} (H_g)$ and $S_{\rm LR} (H_g)$ are statistically indistinguishable.
Based on the results of this (K-S) test, we conclude that there is no significant difference in the absolute magnitude distributions between the ``red'' and ``less red'' JTs for the size range we estimated.
Previously, \cite{wong2014} pointed out that the absolute magnitude distribution differs significantly between red and less red JTs. 
Meanwhile, since the size of JTs in this study was smaller (i.e., fainter) than that in \cite{wong2014}, we think that the difference in the size range of JTs examined may have led to different conclusions.
In other words, the size distribution of small JTs is very similar between the two color groups within the size range that are likely to be collisional fragments or rubble pile.

\section{Discussions} \label{sec:discuss}
In the previous study by \cite{wong2015}, the filters of the Suprime-Cam were changed only once during the four visits, with the cadence of $g^\prime~\rightarrow~g^\prime~\rightarrow~i^\prime~\rightarrow~i^\prime$ (or vice versa). 
The $g^\prime - i^\prime$ distribution obtained from their observations and the $g^\prime - i^\prime$ distribution derived from SDSS-MOC4 are shown in Figure 9 of \cite{wong2015}, and they explain that the bimodality seen in the $g^\prime - i^\prime$ distribution obtained from SDSS-MOC4 for large JTs \citep{wong2014} cannot be confirmed in \cite{wong2015}.
The same trend was observed in \cite{pan2022}. They examined the colors of 573 JTs using Dark Energy Survey (DES) data observed in the $g$, $r$, $i$, and $z$--bands. However, they reported no bimodality in the color distribution of JTs in their sample with absolute magnitudes in the range of 11 $< H <$ 18.
Both \cite{wong2015} and \cite{pan2022} explained that this could be due to the blurring effect of asteroid rotation, because asteroids' rotation contributes to large variance in the color measurements. 
In our observations, the filters were changed in the order $i^\prime~\rightarrow~g^\prime~\rightarrow~i^\prime$ with the time intervals of 20--30~min between the two consecutive exposures. 
Looking at the JT rotation period distribution in \cite{chang2021}, no object with a rotation period of less than two hours has been discovered yet.
Therefore, it is unlikely that the observed JTs brightened and dimmed many times during the above time interval.
Since the $g^\prime - i^\prime$ 
color in this study was calculated by averaging the $i^\prime$-band magnitudes obtained before and after the $g^\prime$-band imaging, the effect of photometric variation with rotation on the $g^\prime - i^\prime$ measurements should be smaller than 
those of previous studies. 
%
Nevertheless, our observations also showed no bimodality in the $g^\prime - i^\prime$ distribution.

Three independent studies (\cite{wong2015}, \cite{pan2022}, and this study) found no bimodality in color distribution for small JTs. This suggests that the absence of bimodality in the color distribution of small JTs may be a genuine phenomenon, although the influence of asteroid rotation cannot be entirely ruled out. 
This may suggest that many of JTs smaller than 10~km such as those observed with the Subaru Telescope have a continuous color distribution between ``red'' (corresponding to D-type asteroids) and ``less red'' (corresponding to P- or C-type asteroids) rather than having a clear distinction between the two. 

To strengthen this possibility, we compared the $g^\prime - i^\prime$ distribution of our sample with that of JTs in the SDSS MOC4 dataset (\url{https://faculty.washington.edu/ivezic/sdssmoc/sdssmoc.html}), which has previously been investigated by \cite{wong2014}, showing a clear bimodality
.
We performed a Kolmogorov-Smirnov (K-S) test with the null hypothesis "our sample has the same distribution as that of the SDSS MOC4 JTs" to determine if the $g^\prime - i^\prime$ distribution of our sample differs from the color distribution of the SDSS MOC4 JTs.
First, we extracted 292 JTs from the SDSS MOC4 catalog using the following conditions: 5.04 $< a$ [au] $<$ 5.4, $e <$ 0.3, and $H <$ 12.3 mag (based on \cite{pan2022}). We designate this group of objects as the SDSS JTs. We randomly selected 44 JTs from the SDSS JTs and compared their cumulative distribution function to that of our unbiased sample of 44 JTs. 
Under the above null hypothesis, the K-S statistic 
from 100,000 two-sample K-S tests was approximately 0.5
. The results show that all the $p$-values are below 0.01.
Therefore, we reject the null hypothesis at the 99\% confidence level. 
This suggests that the color distribution of our sample differs from that of the SDSS JTs, which are considered to have a bimodal color distribution.
%
%
Although the K-S test showed statistical significance, our sample size was small, so further confirmation with larger-scale surveys is expected. When conducting surveys, it would be desirable to take measures to more reliably eliminate the influence of asteroid rotation, such as observing the change in luminosity over at least one rotation period in each band and calculating the average magnitude in each band, before determining the color.

\begin{figure}
\plotone{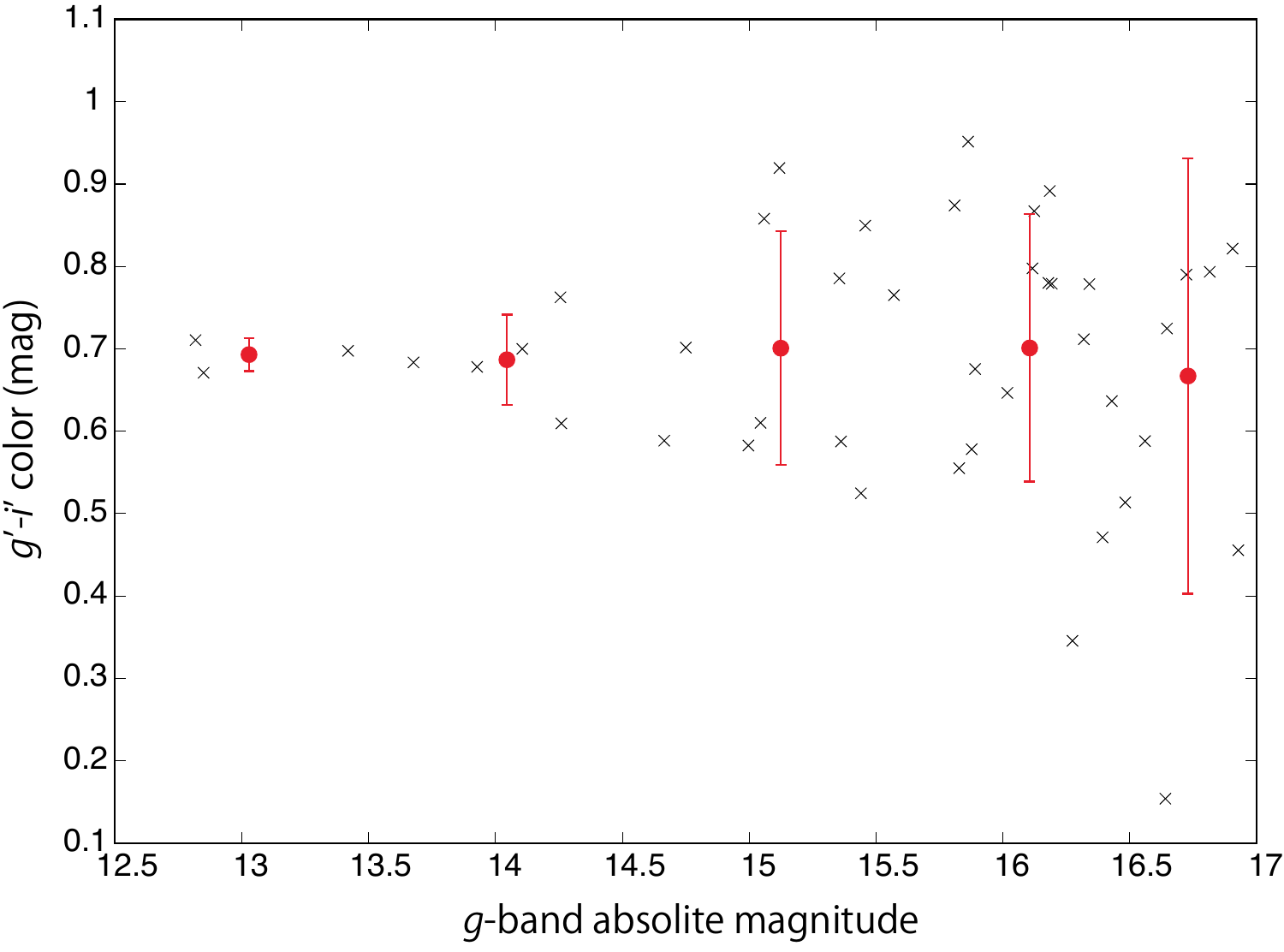}
\caption{Correlation between JT's size and the $g^\prime - i^\prime$ color in our unbiased sample. Circles with error bars represent the average the $g^\prime - i^\prime$ color for each absolute magnitude bin.
\label{fig:size-color}}
\end{figure}

As we mentioned above, the color distribution of large JTs exhibits a bimodality \citep{wong2014},
but such bimodality was not observed in the color distribution of small JTs, with the overall color of small JTs with decreasing size approaching the average color of the less red group of the large JTs \citep{wong2015}.
Figure \ref{fig:size-color} shows the results of examining this trend, as demonstrated by \cite{wong2015}, using our unbiased sample. Within our sample range, the overall average $g^\prime - i^\prime$ color is 0.67 -- 0.70, which largely matches the results of \cite{wong2015} and \cite{pan2022}, though the average $g^\prime - i^\prime$ value in our sample is slightly lower than theirs.

%
Based on these observational results, 
\cite{Wong_2016} 
proposed that less-red and red JTs originated
from the same sources as red and very red KBOs, respectively.
%
When a Trojan asteroid experiences a catastrophic impact, its irradiated surface is lost and volatile interior ices rapidly sublimate, leaving fragments made mainly of rock and water ice. 
These fragments, lacking surface 
volatiles
, become spectrally similar regardless of their origin. Over time, irradiation slightly reddens them, but not as much as if 
volatiles 
were present. 
All collisional fragments, regardless of the color of their progenitor bodies, would eventually attain the same surface color, which would be less-red compared to the Trojans in the red group. 
Thus, collisional evolution reduces red Trojans and increases less-red Trojans, making the population gradually more dominated by less-red objects with decreasing size 
\citep{wong2015}.

On the other hand, we found that the size distribution does not differ between ``red'' and ``less red'' JTs. 
Figure~\ref{fig:csds} shows that the absolute magnitude distributions for both ``red'' and ``less red'' groups follow approximately the absolute magnitude distribution curve obtained by \cite{yoshida2017} with a single filter ($r'$-band).
This means that, within our sample range of 12.9~mag $< H_{g} <$ 16.9~mag, the fractions of ``red'' and ``less red'' objects are independent of their size.
Thus, our results can provide new insights into the collisional evolution of color and size distributions of small JTs.
%
%
If the LUCY mission, first flyby exploration of various types of Jupiter Trojan asteroids \citep{Levison_2021}, reveals the 
local surface colors and/or surface ages inferred from impact craters for D-type and P/C-type JTs, it will definitely provide stronger constraints on their collisional evolution.

\section{Summary} \label{sec:summary}
We used the Subaru Telescope and Suprime-Cam to study the $g^\prime - i^\prime~$ color of the L4 group of Jupiter Trojan asteroids.
Observations were conducted on May 24 and 30, 2017 (UT), covering a total area of 9.2 square degrees. 
The detection limit for this survey was 23.9 mag (in $g^\prime$ band).
To mitigate the effect of light variations due to asteroid rotation, we took images with the $i^\prime~\rightarrow~g^\prime~\rightarrow~i^\prime~$ filter sequence within 40--60~min. 
We obtained the $g^\prime - i^\prime~$ color by calculating the average $i^\prime$-magnitude from the $i^\prime$-band images taken before and after the $g^\prime$-band image.
From the 120 detected JTs, we extracted an unbiased sample of 44 JTs in the range of $R<$5.6~au and $H_{g}<$16.9~mag and derived their $g^\prime - i^\prime~$ color distribution.

The bimodality of "red" and "less red" objects was not evident in the $g^\prime - i^\prime~$ color distribution derived from our sample.
This is in agreement with \cite{wong2015}, who did not find clear bimodality in the color distribution of small JTs also observed by the Subaru Suprime-Cam.
The same trend was reported in \cite{pan2022}, who examined the colors of JTs using Dark Energy Survey data.
This may indicate that many JTs smaller than 10 km, such as those examined in the present work, have a continuous color distribution between the red and less-red groups rather than having a clear distinction between the two groups.
However, this is still highly uncertain because the sample size in the present study is rather small.

%

On the other hand, we also found that there is no difference in the size distribution between ``red'' and ``less red'' JTs within the size range of our sample.
This suggests that in the range of 12.9~mag$<H_{g}<$16.9~mag, ``red'' and ``less red'' JTs are also present in the same proportion, regardless of size.
%
Previous studies reported that small JTs near the faint end observed by the Subaru Suprime-Cam are dominated by less-red objects, which was interpreted as a result of conversion of red objects to less-red fragments by collisional disruption.
Thus, our results can provide new insights into collisional evolution of color and size distribution of small Jupiter Trojan asteroids.
However, the number of samples obtained in our observations this time was small. Therefore, it would be desirable to confirm our results with a larger sample size in the future.

\begin{acknowledgments}
We thank the anonymous reviewer for their constructive suggestions and comments, which greatly strengthened this article.
This work was supported by JSPS KAKENHI Grant No. 20H04617(F.Y.), 24K07123(F.Y.),
22H01286(K.O., T.T., F.Y.), 23K22557(K.O., T.T., F.Y.).
This research is based on data collected at the Subaru Telescope, which is operated by the National Astronomical 
Observatory of Japan. 
We are honored and grateful for the opportunity of observing the Universe from Maunakea, which has the cultural, historical, and natural significance in Hawaii. 
The Pan--STARRS1 Surveys (PS1) and the PS1 public science archive have been made possible through contributions by the 
Institute for Astronomy, the University of Hawaii, the Pan--STARRS Project Office, the Max--Planck Society and its participating institutes, the Max Planck Institute for Astronomy, Heidelberg and the Max Planck Institute for Extraterrestrial Physics, Garching, The Johns Hopkins University, Durham University, the University of Edinburgh, the Queen's University Belfast, the Harvard--Smithsonian Center for Astrophysics, the Las Cumbres Observatory Global Telescope Network Incorporated, the National Central University of Taiwan, the Space Telescope Science Institute, the 
National Aeronautics and Space Administration under Grant No. NNX08AR22G issued through the Planetary Science Division of the NASA Science Mission Directorate, the National Science Foundation Grant No. AST--1238877, the University of Maryland, Eotvos Lorand University (ELTE), the Los Alamos National Laboratory, and the Gordon and Betty Moore Foundation. 

\end{acknowledgments}

\bibliography{fyBib_ver20260130}{}
\bibliographystyle{aasjournal}



\appendix  

\setcounter{section}{0} 
\renewcommand{\thesection}{\Alph{section}} 
\setcounter{equation}{0} 
\renewcommand{\theequation}{\Alph{section}.\arabic{equation}}
\setcounter{figure}{0} 
\renewcommand{\thefigure}{\Alph{section}.\arabic{figure}}
\setcounter{table}{0} 
\renewcommand{\thetable}{\Alph{section}.\arabic{table}}

\section{Observation logs\label{appen:obslog}}

The observation log and other information; the average airmass and seeing size for each field are shown in Table \ref{table:ObsLog}. The survey area located between 12.3 $^{\circ}$ and 15.5 $^{\circ}$ from the L4 point ((RA., DEC.)=(261$^{\circ}$, -23$^{\circ}$)) and 0.2 $^{\circ}$ to 8.7 $^{\circ}$ from the oppositions (RA., DEC.)=(241$^{\circ}$, -21$^{\circ}$) on May 24 (UT) and (RA., DEC.)=(247$^{\circ}$, -22$^{\circ}$) on May 30 (UT).
\begin{longtable}[c]{ccccccccc}

\caption{Observation log 
}
\label{table:ObsLog}
 \\
\hline
 Field & JD     & Filter& RA       & DEC    & Average & Average & Angle & Angle \\
 ID     & mid exp. &         & J.2000 & J2.000 & airmass & seeing   & from   & from   \\
          & (day)      &         & (deg)   & (deg)   &               & (arcsec) & Opp.  (deg)  & L4  (deg) \\
   \hline
   \hline
  \endfirsthead
\multicolumn{9}{l}{\small\slshape continued from previous page} \\
 \hline
 Field & JD      & Filter& RA       & DEC    & Average & Average & Angle & Angle \\
 ID     & mid exp. &         & J.2000 & J2.000 & airmass & seeing   & from   & from   \\
          & (day)      &         & (deg)   & (deg)   &               & (arcsec) & Opp.  (deg)  & L4  (deg) \\
 \hline\hline
 \endhead
 \hline
 \multicolumn{9}{r}{\small\sl table continued on next page}
 \endfoot
\endlastfoot
\hline
F01	&	2457903.941529 	&		$i^\prime$	&	245.612 	&	-21.200 	&	1.370 	&	0.710 	&	1.6 	&	15.5 			\\
	&	2457903.959308 	&		$g^\prime$	&		&		&		&		&		&				\\
	&	2457903.977316 	&		$i^\prime$	&		&		&		&		&		&				\\
F02	&	2457903.942907 	&		$i^\prime$	&	245.612 	&	-21.767 	&	1.385 	&	0.657 	&	1.4 	&	15.4 			\\
	&	2457903.960691 	&		$g^\prime$	&		&		&		&		&		&				\\
	&	2457903.978696 	&		$i^\prime$	&		&		&		&		&		&				\\
F04	&	2457904.007788 	&		$i^\prime$	&	246.091 	&	-19.500 	&	1.726 	&	0.630 	&	2.7 	&	15.3 			\\
	&	2457904.030378 	&		$g^\prime$	&		&		&		&		&		&				\\
	&	2457904.052932 	&		$i^\prime$	&		&		&		&		&		&				\\
F05	&	2457903.954039 	&		$i^\prime$	&	246.091 	&	-20.634 	&	1.387 	&	0.677 	&	1.6 	&	15.1 			\\
	&	2457903.971772 	&		$g^\prime$	&		&		&		&		&		&				\\
	&	2457903.989810 	&		$i^\prime$	&		&		&		&		&		&				\\
F06	&	2457903.945666 	&		$i^\prime$	&	246.091 	&	-21.200 	&	1.376 	&	0.673 	&	1.2 	&	15.0 			\\
	&	2457903.963449 	&		$g^\prime$	&		&		&		&		&		&				\\
	&	2457903.981532 	&		$i^\prime$	&		&		&		&		&		&				\\
F07	&	2457903.944285 	&		$i^\prime$	&	246.091 	&	-21.767 	&	1.385 	&	0.670 	&	0.9 	&	15.0 			\\
	&	2457903.962071 	&		$g^\prime$	&		&		&		&		&		&				\\
	&	2457903.980151 	&		$i^\prime$	&		&		&		&		&		&				\\
F08	&	2457904.005029 	&		$i^\prime$	&	246.566 	&	-18.933 	&	1.669 	&	0.597 	&	3.1 	&	15.0 			\\
	&	2457904.027472 	&		$g^\prime$	&		&		&		&		&		&				\\
	&	2457904.050168 	&		$i^\prime$	&		&		&		&		&		&				\\
F09	&	2457904.006407 	&		$i^\prime$	&	246.566 	&	-19.500 	&	1.698 	&	0.613 	&	2.5 	&	14.9 			\\
	&	2457904.028937 	&		$g^\prime$	&		&		&		&		&		&				\\
	&	2457904.051550 	&		$i^\prime$	&		&		&		&		&		&				\\
F10	&	2457903.991190 	&		$i^\prime$	&	246.566 	&	-20.067 	&	1.582 	&	0.677 	&	2.0 	&	14.7 			\\
	&	2457904.013373 	&		$g^\prime$	&		&		&		&		&		&				\\
	&	2457904.036288 	&		$i^\prime$	&		&		&		&		&		&				\\
F11	&	2457903.952657 	&		$i^\prime$	&	246.566 	&	-20.634 	&	1.379 	&	0.630 	&	1.4 	&	14.6 			\\
	&	2457903.970362 	&		$g^\prime$	&		&		&		&		&		&				\\
	&	2457903.988428 	&		$i^\prime$	&		&		&		&		&		&				\\
F12	&	2457903.947053 	&		$i^\prime$	&	246.566 	&	-21.200 	&	1.376 	&	0.653 	&	0.9 	&	14.5 			\\
	&	2457903.964840 	&		$g^\prime$	&		&		&		&		&		&				\\
	&	2457903.982910 	&		$i^\prime$	&		&		&		&		&		&				\\
F13	&	2457903.948441 	&		$i^\prime$	&	246.566 	&	-21.767 	&	1.392 	&	0.673 	&	0.5 	&	14.5 			\\
	&	2457903.966221 	&		$g^\prime$	&		&		&		&		&		&				\\
	&	2457903.984291 	&		$i^\prime$	&		&		&		&		&		&				\\
F14	&	2457904.003650 	&		$i^\prime$	&	247.045 	&	-18.933 	&	1.644 	&	0.610 	&	3.1 	&	14.5 			\\
	&	2457904.026091 	&		$g^\prime$	&		&		&		&		&		&				\\
	&	2457904.048778 	&		$i^\prime$	&		&		&		&		&		&				\\
F15	&	2457904.002272 	&		$i^\prime$	&	247.045 	&	-19.500 	&	1.647 	&	0.623 	&	2.5 	&	14.4 			\\
	&	2457904.024710 	&		$g^\prime$	&		&		&		&		&		&				\\
	&	2457904.047397 	&		$i^\prime$	&		&		&		&		&		&				\\
F16	&	2457903.992567 	&		$i^\prime$	&	247.045 	&	-20.067 	&	1.582 	&	0.660 	&	1.9 	&	14.3 			\\
	&	2457904.014753 	&		$g^\prime$	&		&		&		&		&		&				\\
	&	2457904.037667 	&		$i^\prime$	&		&		&		&		&		&				\\
F17	&	2457903.993946 	&		$i^\prime$	&	247.045 	&	-20.633 	&	1.608 	&	0.673 	&	1.4 	&	14.2 			\\
	&	2457904.016132 	&		$g^\prime$	&		&		&		&		&		&				\\
	&	2457904.039045 	&		$i^\prime$	&		&		&		&		&		&				\\
F18	&	2457903.951202 	&		$i^\prime$	&	247.045 	&	-21.200 	&	1.384 	&	0.650 	&	0.8 	&	14.1 			\\
	&	2457903.968982 	&		$g^\prime$	&		&		&		&		&		&				\\
	&	2457903.987049 	&		$i^\prime$	&		&		&		&		&		&				\\
F19	&	2457903.949821 	&		$i^\prime$	&	247.045 	&	-21.767 	&	1.392 	&	0.683 	&	0.2 	&	14.0 			\\
	&	2457903.967602 	&		$g^\prime$	&		&		&		&		&		&				\\
	&	2457903.985672 	&		$i^\prime$	&		&		&		&		&		&				\\
F20	&	2457898.039337 	&		$i^\prime$	&	247.520 	&	-18.450 	&	1.758 	&	0.893 	&	7.0 	&	14.2 			\\
	&	2457898.058637 	&		$g^\prime$	&		&		&		&		&		&				\\
	&	2457898.075216 	&		$i^\prime$	&		&		&		&		&		&				\\
F21	&	2457898.040723 	&		$i^\prime$	&	247.520 	&	-19.017 	&	1.792 	&	0.967 	&	6.8 	&	14.1 			\\
	&	2457898.060035 	&		$g^\prime$	&		&		&		&		&		&				\\
	&	2457898.076673 	&		$i^\prime$	&		&		&		&		&		&				\\
F22	&	2457898.030598 	&		$i^\prime$	&	247.520 	&	-19.583 	&	1.699 	&	0.990 	&	6.7 	&	13.9 			\\
	&	2457898.049867 	&		$g^\prime$	&		&		&		&		&		&				\\
	&	2457898.066615 	&		$i^\prime$	&		&		&		&		&		&				\\
F23	&	2457904.000861 	&		$i^\prime$	&	247.520 	&	-20.150 	&	1.640 	&	0.633 	&	1.9 	&	13.8 			\\
	&	2457904.023036 	&		$g^\prime$	&		&		&		&		&		&				\\
	&	2457904.046019 	&		$i^\prime$	&		&		&		&		&		&				\\
F24	&	2457903.995329 	&		$i^\prime$	&	247.520 	&	-20.717 	&	1.610 	&	0.663 	&	1.4 	&	13.7 			\\
	&	2457904.017509 	&		$g^\prime$	&		&		&		&		&		&				\\
	&	2457904.040423 	&		$i^\prime$	&		&		&		&		&		&				\\
F25	&	2457903.996712 	&		$i^\prime$	&	247.520 	&	-21.283 	&	1.637 	&	0.670 	&	0.9 	&	13.6 			\\
	&	2457904.018892 	&		$g^\prime$	&		&		&		&		&		&				\\
	&	2457904.041803 	&		$i^\prime$	&		&		&		&		&		&				\\
F26	&	2457898.081051 	&		$i^\prime$	&	248.000 	&	-18.533 	&	2.363 	&	0.740 	&	7.4 	&	13.7 			\\
	&	2457898.094829 	&		$g^\prime$	&		&		&		&		&		&				\\
	&	2457898.108859 	&		$i^\prime$	&		&		&		&		&		&				\\
F27	&	2457898.037877 	&		$i^\prime$	&	248.000 	&	-19.100 	&	1.746 	&	0.987 	&	7.3 	&	13.6 			\\
	&	2457898.057138 	&		$g^\prime$	&		&		&		&		&		&				\\
	&	2457898.073791 	&		$i^\prime$	&		&		&		&		&		&				\\
F28	&	2457898.032066 	&		$i^\prime$	&	248.000 	&	-19.667 	&	1.702 	&	0.850 	&	7.1 	&	13.4 	\\
	&	2457898.051332 	&		$g^\prime$	&		&		&		&		&		&				\\
	&	2457898.068016 	&		$i^\prime$	&		&		&		&		&		&				\\
F29	&	2457903.999476 	&		$i^\prime$	&	248.000 	&	-20.233 	&	1.620 	&	0.657 	&	2.0 	&	13.3 			\\
	&	2457904.021659 	&		$g^\prime$	&		&		&		&		&		&				\\
	&	2457904.044606 	&		$i^\prime$	&		&		&		&		&		&				\\
F30	&	2457903.998095 	&		$i^\prime$	&	248.000 	&	-20.800 	&	1.624 	&	0.663 	&	1.6 	&	13.2 			\\
	&	2457904.020278 	&		$g^\prime$	&		&		&		&		&		&				\\
	&	2457904.043185 	&		$i^\prime$	&		&		&		&		&		&				\\
F31	&	2457898.082433 	&		$i^\prime$	&	248.479 	&	-18.617 	&	2.369 	&	0.743 	&	7.8 	&	13.3 			\\
	&	2457898.096212 	&		$g^\prime$	&		&		&		&		&		&				\\
	&	2457898.110314 	&		$i^\prime$	&		&		&		&		&		&				\\
F32	&	2457898.079562 	&		$i^\prime$	&	248.479 	&	-19.183 	&	2.324 	&	0.793 	&	7.7 	&	13.1 			\\
	&	2457898.093371 	&		$g^\prime$	&		&		&		&		&		&				\\
	&	2457898.107361 	&		$i^\prime$	&		&		&		&		&		&				\\
F33	&	2457898.036411 	&		$i^\prime$	&	248.479 	&	-19.750 	&	1.735 	&	0.950 	&	7.6 	&	12.9 			\\
	&	2457898.055635 	&		$g^\prime$	&		&		&		&		&		&				\\
	&	2457898.072298 	&		$i^\prime$	&		&		&		&		&		&				\\
F34	&	2457898.033497 	&		$i^\prime$	&	248.479 	&	-20.317 	&	1.723 	&	0.827 	&	7.5 	&	12.8 			\\
	&	2457898.052757 	&		$g^\prime$	&		&		&		&		&		&				\\
	&	2457898.069511 	&		$i^\prime$	&		&		&		&		&		&				\\
F35	&	2457898.083817 	&		$i^\prime$	&	248.954 	&	-18.700 	&	2.376 	&	0.747 	&	8.3 	&	12.8 			\\
	&	2457898.097593 	&		$g^\prime$	&		&		&		&		&		&				\\
	&	2457898.111770 	&		$i^\prime$	&		&		&		&		&		&				\\
F36	&	2457898.078106 	&		$i^\prime$	&	248.954 	&	-19.267 	&	2.263 	&	0.783 	&	8.1 	&	12.6 			\\
	&	2457898.091903 	&		$g^\prime$	&		&		&		&		&		&				\\
	&	2457898.105902 	&		$i^\prime$	&		&		&		&		&		&				\\
F37	&	2457898.034953 	&		$i^\prime$	&	248.954 	&	-19.833 	&	1.710 	&	0.933 	&	8.0 	&	12.5 			\\
	&	2457898.054248 	&		$g^\prime$	&		&		&		&		&		&				\\
	&	2457898.070919 	&		$i^\prime$	&		&		&		&		&		&				\\
F38	&	2457898.085205 	&		$i^\prime$	&	249.433 	&	-18.783 	&	2.382 	&	0.727 	&	8.7 	&	12.3 			\\
	&	2457898.098973 	&		$g^\prime$	&		&		&		&		&		&				\\
	&	2457898.113226 	&		$i^\prime$	&		&		&		&		&		&				\\
\hline
\end{longtable}
\section{Example images for the visual inspection of JTs\label{appen:ex_visualinspection}}

The following diagram was created for each JT candidate to facilitate visual inspection. The position of the JT candidate on the image at each visit is indicated by a square, together with a magnified image within the square. If an object is present in a square, the JT candidate is considered real.

\begin{figure}
\plotone{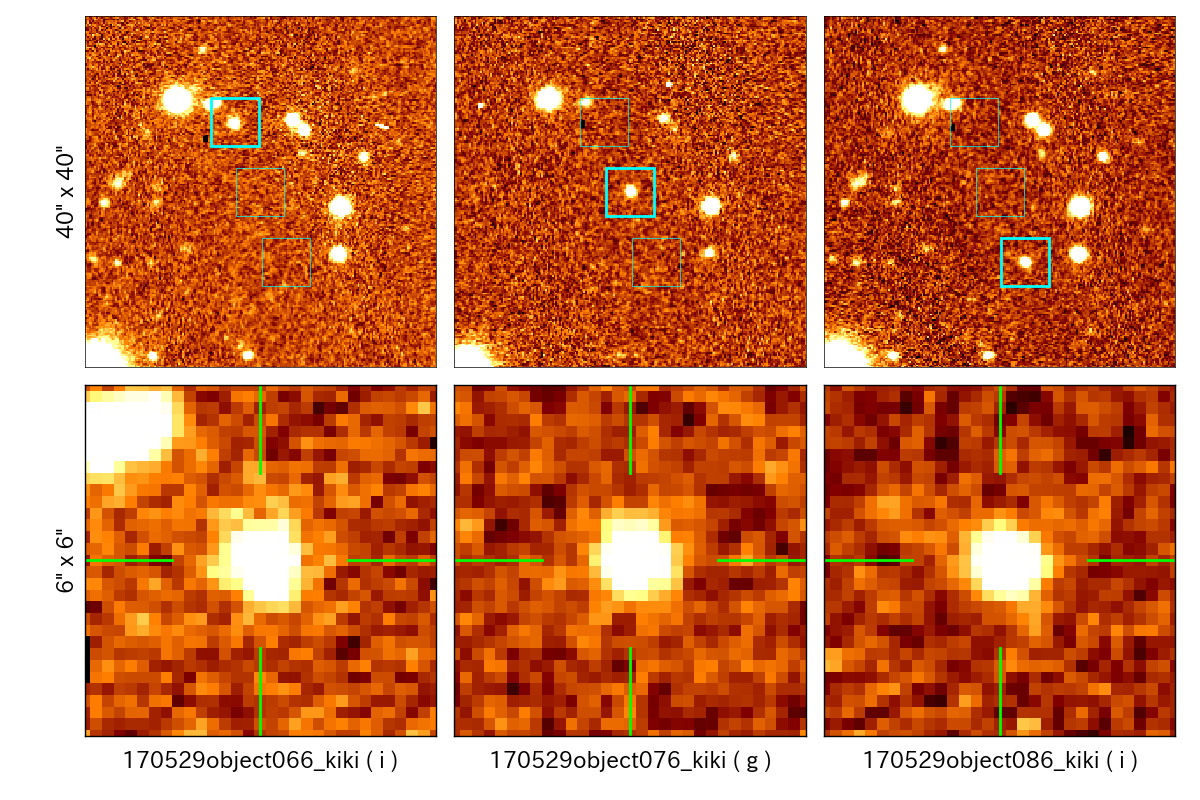}
\caption{An example of the diagram used for visual inspection. 
The top three panels show squares representing the position of each of the three visits, from left to right of the moving object corresponding to the JT motion derived from the light source catalog.
The images at the first and third visits are taken with the $i^\prime$-band filter, while the $g^\prime$-band filter was used for the second visits. 
The bottom panels are an enlarged version of the image within the upper square with the thick line. 
The field of view is 6” $\times$ 6". 
The green lines are drawn to make it easier to see the center of the square. \label{eye-inspection}}
\end{figure}
\end{document}